\documentclass[preprint,aps]{revtex4}
\usepackage{graphicx}
\usepackage{psfrag}

\begin{document}
\bibliographystyle{prsty}

\title{PLASTIC RESPONSE OF A 2D AMORPHOUS SOLID TO QUASI-STATIC SHEAR :\\
II - DYNAMICAL NOISE AND AVALANCHES IN A MEAN FIELD MODEL}

\author{Ana\"el Lema\^{\i}tre$^{(1)}$}
\author{Christiane Caroli$^{(2)}$}
\affiliation{$^{(1)}$ Institut Navier -- LMSGC, CNRS, UMR 113, 2 all\'ee K\'epler,77420 Champs-sur-Marne, France}
\affiliation{$^{(2)}$ INSP, Universit\'e Pierre et Marie Curie-Paris 6, 
CNRS, UMR 7588, 140 rue de Lourmel, 75015 Paris, France}

\date{\today}

\begin{abstract}
We build a minimal, mean-field, model of plasticity of amorphous
solids, based upon a phenomenology of dissipative events derived, in
a preceding paper [A. Lemaitre, C. Caroli, arXiv:0705.0823] 
from extensive molecular simulations. It reduces to the dynamics of an
ensemble of identical shear transformation zones interacting via the
dynamic noise due to the long ranged elastic fields induced by zone
flips themselves. We find that these ingredients are sufficient to
generate flip avalanches with a power-law scaling with system size,
analogous to that observed in molecular simulations. We further show
that the scaling properties of avalanches sensitively depend on the
detailed shape of the noise spectrum. This points out the importance
of developing a realistic coarse-grained description of elasticity in
these systems.
\end{abstract}

\maketitle
\section{Introduction}
Much effort has recently been devoted to the development of
theoretical descriptions of plasticity of amorphous media. They aim at
proposing macroscopic constitutive laws consistent with the microscopic
information emerging from a wealth of numerical results accumulated
over the past twenty years. In particular, it is now well established
that plastic deformation in these systems proceeds via irreversible
sudden
rearrangements of small clusters of atoms. Even at zero
temperature, these flips occur intermittenly. These empirical facts
are the basis of two recently proposed phenomenologies: the STZ
(shear transformation zone) theory~\cite{FalkLanger1998} and SGR (soft glass rheology) model~\cite{Sollich1998}.
Both represent the sheared disordered solid as a set of spatially
random and {\it independent} "zones" or "traps" of small size
embedded in a homogeneous background. These structures are metastable,
so that, when loaded elastically by the external driving strain, in
the absence of noise elastic loading would stop at an instability
threshold where they flip into an unstressed state. Moreover, in both
models, structural disorder gives rise to a noise acting in parallel
with advective loading, thus resulting in intermittent flips occurring
before absolute instability is reached. An important point is that
they introduce noise via Arrhenius factors associated with a
constant, {\it strain-rate independent effective temperature}.

In an attempt at testing the validity of the assumption concerning
flip independence, Maloney and Lema\^{\i}tre~\cite{MaloneyLemaitre2006} 
later carried out extensive
numerical simulations on 2D glasses of various sizes $L \times L$ in the
athermal  quasi-static regime ($T = 0$; vanishing strain rate
$\dot{\gamma} \rightarrow 0$), hereafter abbreviated as AQS. In this
regime, pure elastic, reversible, loading is interrupted by randomly
spaced discontinuous stress drops associated with the
(quasi-instantaneous) plastic events. They found that these events
can be interpreted as avalanches involving a varying number $n$ of
elementary zone flips. In the stationary state, the avalanche size
$n$, hence the stress drop amplitude $\Delta\sigma$ are broadly
distributed, and their averages are {\it system-size dependent}. The
average avalanche size scales roughly as $\langle n \rangle \sim L$~\cite{MaloneyLemaitre2004}. 
Distributed
avalanches have also been found by Bailey et al in 3D simulations~\cite{BaileySchiotzLemaitreJacobsen2007},
with an approximate scaling $\langle n \rangle \sim L^{3/2}$.

In the 2D simulations, flips show up clearly as elastic quadrupolar
structures in the energy and the non-affine displacement field,
consistent with their representation as shear transformations of
Eshelby-like inclusions~\cite{Eshelby1957}. Avalanches are known to result from
long range interactions. In the present case interzone elastic couplings
mediated by the background medium are thus likely to be responsible for
the avalanche behavior.

Motivated by this analysis, we have performed in a previous, companion,
paper~\cite{LemaitreCaroli2007} an extensive numerical study of the
evolution with strain of the non-affine field in a 2D LJ glass. It
substantiates the importance of interzone elastic couplings and leads
to the following picture,
consistent with that proposed already long ago by Argon {\it et al}~\cite{ArgonBulatovMottSuter1995}:
via its associated quadrupolar
field, a zone flip induces, at any other zone site in the system, a
shift of the strain level whose amplitude and sign depend on the
relative position of target and source. These shifts may bring some
zones past their instability threshold, hence triggering an
avalanche. For the zones which do not take part in an avalanche, the
flip-induced elastic signals act as an {\it intrinsic dynamical noise}
the frequency of which scales as  the strain rate $\dot{\gamma}$. The
statistical study of particle motion shows that this noise dominates
largely over fluctuations associated with non affine deformations
during the elastic episodes separating the plastic events.

In this article we propose a minimal model which incorporates as
simply as possible the essential features of this phenomenology in
terms of a set of identical spatially random zones embedded in a 2D
elastic continuum, driven by external loading toward their instability
threshold and coupled via flip-induced quadrupolar elastic fields.

In Section II we define our model in detail. We discuss the various
relevant time scales in steady plastic flow, which allows us to
identify, for finite systems, a size-dependent quasi-static regime
where avalanches can be considered instantaneous. We then study
numerically, in a mean-field approximation for the elastic noise, the
steady state dynamics for different system sizes. We find that
dissipation occurs via broadly distributed flip avalanches, the average size of
which exhibits a power law scaling with system size: $<n> \sim
L^{\beta}$. However, the exponent, $\beta \simeq 0.3$, is definitely
smaller than the value, of order $1$, measured in reference~\cite{MaloneyLemaitre2004}.

So, in Section III we discuss the various simplifications involved in
our modelling of zones and of their elastic field. This leads us to
test a second version of the model in which we assume
empirically a gaussian spectrum for the elastic noise. While
preserving the existence of the avalanche dynamics, this modified
model turns out to predict the correct power law scaling for $<n>$. We
further show that, in this case, an analytical estimate based on a
Fokker-Planck-like approximation yields the same prediction for the scaling
exponent.

Although still preliminary, these results lead us to conclude that
phenomenologies of plasticity of amorphous media should definitely
incorporate into their basic ingredients the
strain-rate dependent elastic noise generated by the zone flips
themselves.

\section{Elastically coupled zones and avalanches:}

\subsection{The model}

We consider an ensemble of $N$ identical zones of size $a$, shear
elastic modulus $\mu$, randomly distributed with a fixed density
$\rho$ in a $2D$ elastic medium of lateral size $L$. The average
distance between nearest zones $d$ is fixed and given by :
$d^{2} = L^{2}/N = a^{2}/\rho$. The elastic state of each zone is
characterized by an internal strain $\epsilon_{i}$, which measures
the departure from its zero stress state. The $\epsilon_{i}$'s lie
below a common instability threshold $\epsilon_{c}$. The athermal system is
driven by external shear at rate $\dot{\epsilon}$, which advects all
$\epsilon_{i}$'s. When a zone reaches  $\epsilon_{c}$, it disappears
while releasing an amount of internal strain $\Delta\epsilon_{0}$. At
the same time another one is created, at an uncorrelated position,
with zero initial stress (hence zero internal strain).

\begin{figure}
\begin{picture}(40,40)(-20,-20)
\put(-35,0){\line(1,0){70}}
\put(0,-35){\line(0,1){70}}
\thicklines
\put(20,0){\vector(0,1){15}}
\put(-20,0){\vector(0,-1){15}}
\put(0,20){\vector(1,0){15}}
\put(0,-20){\vector(-1,0){15}}
\end{picture}
\caption{\label{fig:dipole}
The perturbation due to a localized plastic event corresponds to the elastic response
to two force dipoles.}
\end{figure}
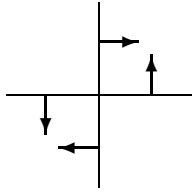

During a zone flip the cluster of atoms forming the zone jump into a
configuration compatible with the externally imposed strain. This
process relaxes the intra-zone stress and, at the same time, deforms the
surrounding elastic medium. Following Picard {\it et al}~\cite{PicardAjdariLequeuxBocquet2004}, its field can be
represented as due to two force dipoles (Figure \ref{fig:dipole}).
We take for this
strain field its expression in an infinite medium. At relative position
${\bf r} = (r, \theta)$ from the flipping center:
\begin{equation}
\label{eq:field}
\Delta\epsilon ({\bf r}) = \frac{2}{\pi}
\frac{a^{2}\Delta\epsilon_{0}} {r^{2}} \cos 4\theta
\end{equation}
where $\theta$ is measured from the shearing direction.

It has quadrupolar symmetry, hence zero average.
$\Delta\epsilon_{0}$ and $\epsilon_{c}$ can be related using the
following argument: the amount of stress $\mu \Delta\epsilon_{0}$
released inside the zone by a flip gives rise to a spatially
averaged, macroscopic stress drop $\Delta\bar{\sigma_{0}} = \kappa
\frac{a^{2}}{L^{2}} \mu \Delta\epsilon_{0}$, where the number
$\kappa = \mathcal{O}(1)$ depends on the shape of the system. From
now on, we assume that $\kappa = 1$. In stationary state, over a
large strain interval, the average number of zone flips is
$N \Delta\epsilon/\epsilon_{c}$. The associated macroscopic plastic
stress release $N \Delta\bar{\sigma_{0}}$ must balance the increase of
elastic stress, $\mu \Delta\epsilon$, due to loading. Hence
\begin{equation}
\label{eq:epsilonzero}
\frac{\Delta\epsilon_{0}}{\epsilon_{c}} = \frac{L^{2}}{N a^{2}} =
\frac{d^{2}}{a^{2}}
\end{equation}

The duration $\tau_{0}$ of a flip is controlled by the time necessary
to radiate elastic energy out of the zone region, i.e. by radiative
acoustic damping, so $\tau_{0} \sim a/c_{s}$, with $c_{s}$ a sound
speed. This acoustic signal, emitted from site ${\bf r}_{i}$,
propagates throughout and modifies the strains of all other zones $j$
which adjust, over a time $\sim \tau_{0}$, to this space-dependent
shift $\Delta\epsilon ({\bf r}_{ij})$. Flip signals thus constitute a
noise acting on the $\epsilon_{i}$'s.

In steady state, the average flip rate in the whole system is
\begin{equation}
\label{eq:rate}
\mathcal{R}_{flip} = \delta t_{flip}^{-1} = N
\frac{\dot{\epsilon}}{\epsilon_{c}} = \frac{L^{2}}{d^{2}}
\frac{\dot{\epsilon}}{\epsilon_{c}}
\end{equation}
If flips occur independently, i.e. in the absence of avalanches, the
noise correlation time is $\tau_{0}$, and the QS regime, where flips
can be assumed instantaneous, corresponds to $\delta\tau_{flip} >>
\tau_{0}$, that is to:
\begin{equation}
\label{eq:QS1}
\dot{\epsilon} << \dot{\epsilon}_{flip} = 
\frac{c_s}{a}\,\epsilon_c\,\frac{d^2}{L^2}
\end{equation}
For a glass-like system, with $\epsilon_{c} \sim 1\% $, $a \sim 1$nm,
zone density $a^{2}/d^{2} \sim 10^{-1}$, lateral size $L \sim 1$mm,
this yields the loose criterion $\dot{\epsilon} << 1$.

Now, a first elastic noise signal may drive some $\epsilon_{j}$'s
beyond $\epsilon_{c}$, hence trigger secondary flips, thus initiating
an avalanche whose duration $\tau_{av}$ is set by sound propagation.
For a very conservative estimate, we take the average distance between
  successive flips to be $L$. This leads to a duration $\tau_{av}
= <n>L/c_{s}$, with $<n>$ the average avalanche size. It must be
compared with the average time interval~\cite{footnote}
between avalanches $\delta t_{av}$, given by:
\begin{equation}
\label{eq:tav}
\delta t_{av}^{-1} = \frac{N \dot{\epsilon}}{<n> \epsilon_{c}}
\end{equation}
The QS condition then becomes
\begin{equation}
\label{eq:QS2}
\dot{\epsilon} << \frac{\epsilon_{c} c_{s}}{N L} = \frac{a}{L}
\dot{\epsilon}_{flip}
\end{equation}

Let us emphasize here an important point, usually ignored in related earthquake
models centered on the issue of criticallity:
the quasi-static range is limited by the acoustic delays
controlling avalanche spreading, hence shrinks with system size.

We will assume from now on that the QS condition is fulfilled. We
thus model the noise by instantaneous shifts $\delta\epsilon_{i}$ of
the zone strains. Moreover, we treat our model, now coined
"E model", in the mean-field
approximation, i.e. assume that the $\delta\epsilon_{i}$'s are
independent random variables, which amounts to neglecting space
correlations between flip centers. That is, we take the spectrum
$\Pi_{E}(\delta\epsilon)$ of these noise signals to be that due to a
spatially uniform distributions of sources truncated at the average
distance between nearest zones $d$.
\begin{equation}
\label{eq:PiE}
\Pi_{E} (\delta\epsilon) = \frac{1}{\pi \left(L^{2} - d^{2}\right)}
\int_{d}^{L} r dr \int_{-\pi}^{\pi} d\theta\,\delta\left(
\Delta\epsilon(r) - \delta\epsilon\right)
\end{equation}
with $\Delta\epsilon(r)$ given by equation~(\ref{eq:field}).

\begin{figure}
\psfrag{pi}{{\large $\Pi_{E}$}}
\psfrag{eps}{{$\delta\epsilon/\epsilon_c$}}
\includegraphics[width=0.5\textwidth,clip]{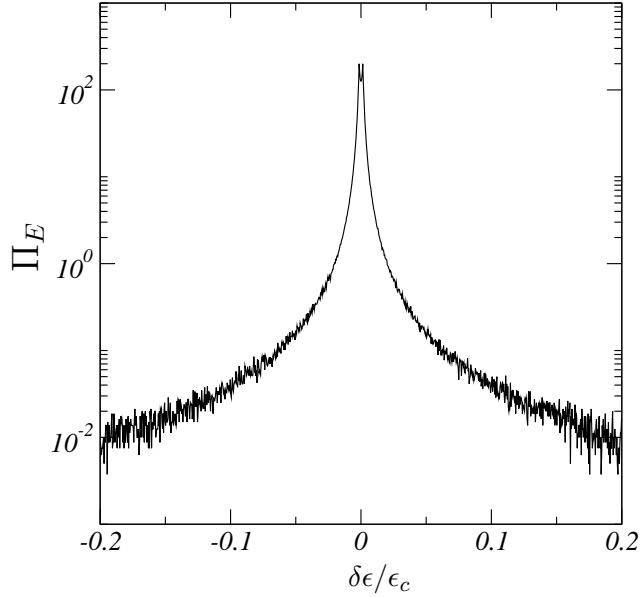}
\caption{\label{fig:spectrum}
$\Pi_{E}$, for $N=500$.
}
\end{figure}

$\Pi_{E}(\delta\epsilon)$, plotted on Figure \ref{fig:spectrum}, is
size-dependent, and has variance
\begin{equation}
\label{eq:variance}
M_{2}(N) = \frac{2\epsilon_{c}^{2}}{\pi^{2} N}
\end{equation}
It exhibits a narrowly peaked structure, associated with distant
zones, and vanishes beyond the cut-off $(2a^{2}
\Delta\epsilon_{0}/\pi d^{2})$, so that all its moments are finite.
Yet, it presents broad, power law, tails. For example, its first half
moment
\begin{equation}
\label{eq:M1}
M_{1+}^{E} =
\int_{0}^{\infty}\,d(\delta\epsilon)\,\delta\epsilon\,\Pi_{E}(\delta\epsilon) =
\frac{8\epsilon_{c}}{\pi}\,\frac{\ln N}{N}
\end{equation}

At this stage, the model can be summarized into the following two-step
algorithm. Since, in the QS regime, the dynamics between instantaneous
avalanches reduces to steady drift of all $\epsilon_{i}$'s at constant
loading rate:

{\it (i)} Starting from an initial configuration where all
$\epsilon_{i} < \epsilon_{c}$, we first identify $\epsilon_{M} = Max
(\epsilon_{i})$. We shift all $\epsilon_{i}$ by
$\Delta\epsilon_{drift} = \epsilon_{c} - \epsilon_{M}$, and the first
avalanche is triggered. Then:

{\it (ii)} Zone $M$ flips, i.e. is removed, while a new one is
introduced at zero strain. This first flip emits a noise signal which
randomly shifts all the other zones: $\epsilon_{i} \rightarrow
\epsilon_{i1} = \epsilon_{i} + \delta\epsilon_{i}$, where the
$\delta\epsilon_{i}$'s are independent and distributed according to
$\Pi_{E}(\delta\epsilon)$. If all $\epsilon_{i1} < \epsilon_{c}$, we
are back to step {\it (i)}. Otherwise, an avalanche starts: we count
the number $q_{1}$ of zones which flip at this stage and are replaced
by new, unstrained, ones.

{\it (iii)} Each zone must now receive $zq_{1}$ signals, which we
treat successively. The first of these yields a new shifted
configuration $\{\epsilon_{i2}\}$ which produces $q_{2}$ new flips.
$z$ is then updated to $z \rightarrow z -1 + q_{1}$, etc\ldots The
avalanche stops when $z$ vanishes. Its size is $n = 1 + \sum q_{i}$.

\subsection{Numerical results}

In order to study the statistical properties of our model in steady
state, we eliminate the initial transients ($\epsilon/\epsilon_{c} <
2$) and perform ergodic averaging over long strain intervals involving
$\gtrsim 10^{5}$ avalanches.

\begin{figure}
\psfrag{pdf}{{\large $\varpi(n)$}}
\includegraphics[width=0.5\textwidth,clip]{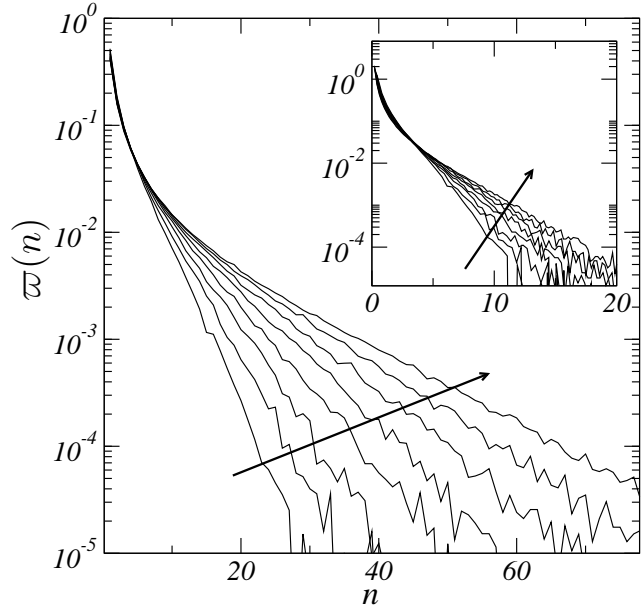}
\caption{\label{fig:piE}
Distribution $\varpi(n)$ of avalanche sizes, for our elastic model,
and for all system sizes. 
Insert: $\langle n\rangle\,\varpi(n)$ vs $n/\langle n\rangle$.
}
\end{figure}

\begin{figure}[b]
\psfrag{nav}{{\large $\log\langle n\rangle$}}
\psfrag{sys_size}{{\large $\log N$}}
\includegraphics[width=0.45\textwidth,clip]{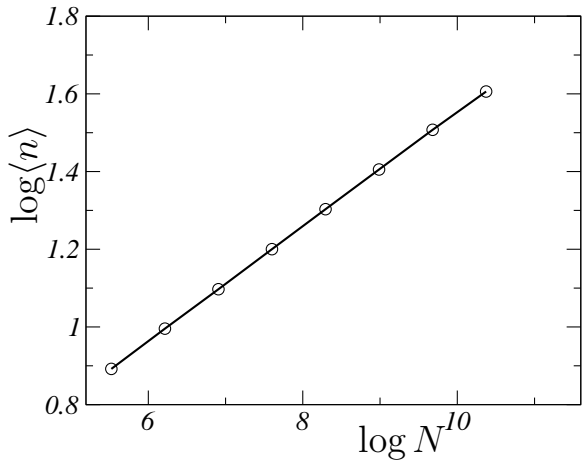}
\caption{\label{fig:sizeE}
Average avalanche size $\langle n\rangle$ vs $N$ for the E model.
}
\end{figure}

As shown on Figure \ref{fig:piE}, their size distribution
$\varpi_{E}(n)$ depends only weakly on the size $N$ for small $n$(
$\lesssim 5$). For $n \gtrsim 5$, it presents a quasi-exponential
tail, which broadens noticeably with increasing $N$. This size
dependence reflects into the growth with $N$
of the average avalanche size (see Figure \ref{fig:sizeE}) which we
find to fit closely, over the whole $N$-range (two decades) a power
law behavior:
\begin{equation}
\label{eq:scalingE}
\langle n \rangle \sim N^{\alpha_{E}} \,\,\,\,\,\,\,\,\,\, \alpha_{E} =
0.147
\end{equation}

When plotting $\langle n \rangle\varpi_{E}(n)$ versus $n/\langle n
\rangle$ (see insert of Figure \ref{fig:piE}) it turns out, however,
that $\varpi_{E}(n)$ does not obey a simple scaling. Even though power
law like decay may be identified on a limited, small-$n$, range, this
by no means allows us to conclude to self-criticallity -- at variance
with a previous claim by Chen et al on a related earthquake model~\cite{ChenBakObukhov1991}.

We must now confront the above results with those of the molecular
simulations. Clearly, our highly simplified model shows that long range
interzone elastic couplings indeed produce broadly distributed
avalanches with an average size growing as a power law of system size.
However, at this stage, agreement is merely qualitative, since
{\it (i)} The scaling exponent $\alpha_{E}$ differs from the
simulation value $\alpha_{sim} \simeq 1/2$.
{\it (ii)} In the molecular simulations~\cite{MaloneyLemaitre2006}, rescaling avalanche sizes by
their $N$-dependent average results in a rather good collapse of
data, which does not hold in the model.

\section{Discussion}

\subsection{Approximations of the E model}
The above comparison leads us to discuss in more detail the
assumptions involved in our minimal model.

First of all, let us consider more closely our representation of
elastic couplings in our finite system. When approximating the static
elastic propagator by its value for an infinite medium, we neglect
the finite global elastic recoil which necessarily accompanies, when
driving at imposed strain, the macroscopic stress release after each
flip. This recoil, of order $1/L^{2}$, should be modelled as a common
backward shift, $(-\xi/N)$ which should be added to the noise
$\delta\epsilon_{i}$. We have rerun the E model for $\xi = 1$ and $2$.
We find that a finite recoil leaves the power law scaling of $\langle
n\rangle$ (equation (\ref{eq:scalingE})) unchanged, the intuitively
expected avalanche size reduction showing up only in a slow decrease
with $\xi$ of the prefactor.

Let us now try to list the various simplifications underlying our
representation of the coupled zones. They can be separately into
approximations concerning respectively (a) the zones themselves and
(b) the elasticity of the embedding medium.

(a) {\it Zones:}

We have taken them to be identical, i.e. to have the
same threshold strain $\epsilon_{c}$, and the same shear modulus
$\mu$, which we have assumed to be constant up to $\epsilon_{i} =
\epsilon_{c}$. However, clearly, in atomically disordered systems,
both $\mu$ and $\epsilon_{c}$ depend on details of the internal
structure of the zone and of its immediate vicinity. So, these two
parameters are certainly distributed about characteristic averages.
For example, a signature of the spread of $\mu$ is the observation in
the LJ simulations~\cite{LemaitreCaroli2007} of instances in which a zone "overtakes"
another one during an elastic loading episode. Moreover, we know~\cite{MaloneyLemaitre2004a}
  and
have checked in ref.~\cite{LemaitreCaroli2007} that significant elastic softening occurs near the
threshold, where a metastability barrier vanishes.

On the other hand, it was observed in~\cite{LemaitreCaroli2007}
that, frequently, a first
flip does not result in the disappearance of the zone. Rather, this
persists after a finite strain release, several flips being needed for
it to finally "die out". Describing this behavior would demand a
multistate zone model.

(b) {\it Elastic couplings:}

We have represented the embedding elastic medium as a homogeneous
continuum, and have taken for the elastic field (equation
(\ref{eq:field})) its expression for an infinite sytem. We believe that
this last approximation does not affect avalanche size scaling.
Indeed,  an elementary dimensional analysis for the stain field
$\Delta\epsilon$ in a finite $L \times L$ box shows that all its
moments retain the same $L$ dependence as those of the $\Pi_{E}$
spectrum (equation (\ref{eq:PiE})) used in Section~II.

Note however that the homogeneous continuum approximation itself is probably
overschematic, in particular at short distances, where numerical
studies of the response to a localized force have shown that it is
dominated by disorder-induced fluctuations~\cite{LeonforteTanguyWittmerBarrat2004}. 
This certainly contributes to a decrease of the tail of noise spectrum.

Pending more quantitative information about these finer disorder
effects, we now propose to test the robustness of our minimal model by
investigating the avalanche statistics under the empirical assumption of a
gaussian elastic noise spectrum.

\subsection {Avalanches in a gaussian noise model}

In this "G model", we choose the gaussian noise spectrum
$\Pi_{G}(\delta\epsilon)$ to have the same variance $ M_{2}(N)$ as
that (equation (\ref{eq:variance})) for the E model. The algorithm is
then implemented as already described.

\begin{figure}[b]
\psfrag{nav}{{\large $\log\langle n\rangle$}}
\psfrag{sys_size}{{\large $\log N$}}
\includegraphics[width=0.45\textwidth,clip]{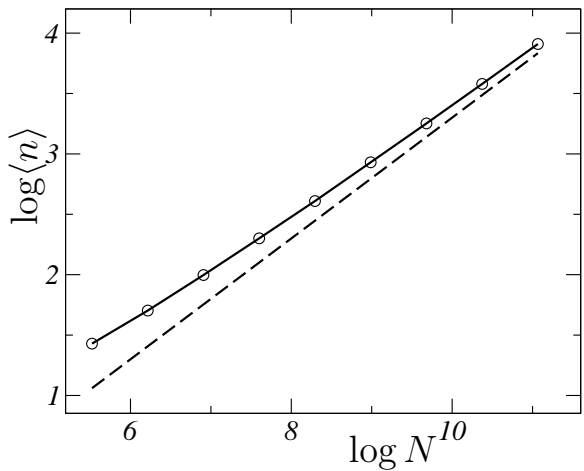}
\caption{\label{fig:sizeG}
Average avalanche size $\langle n\rangle$ vs $N$ for model G.
The dashed line has slope $1/2$.
}
\end{figure}

Here again, we find broadly distributed avalanches, whose average size
increases with $N$ The data, shown on Figure \ref{fig:sizeG}, are
consistent with the asymptotic (see insert) power-law behavior:
\begin{equation}
\label{eq:scalingG}
\langle n \rangle \sim N^{\alpha_{G}} \,\,\,\,\,\,\,\,\,\, \alpha_{G} =
1/2
\end{equation}

\begin{figure}
\psfrag{pdf}{{\large $\varpi(n)$}}
\includegraphics[width=0.5\textwidth,clip]{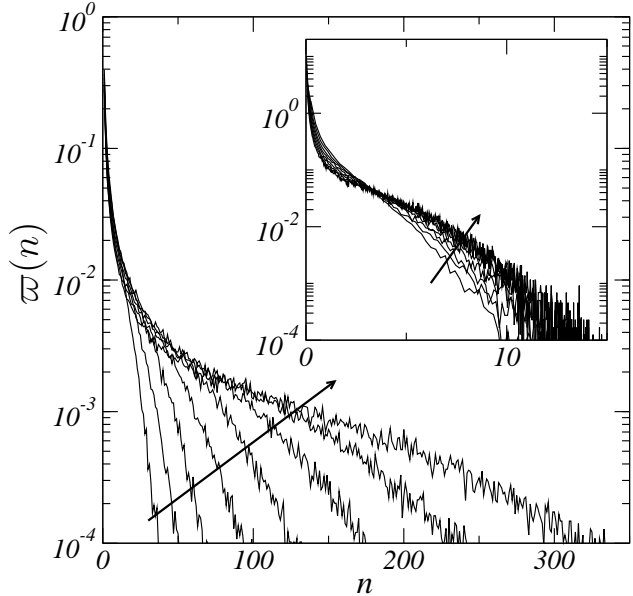}
\caption{\label{fig:piG} 
Distribution $\varpi(n)$ of avalanche sizes, for model G and all system sizes. 
Inserts: $\langle n\rangle\,\varpi(n)$ vs $n/\langle n\rangle$.
}
\end{figure}

Moreover, Figure \ref{fig:piG} shows that the rescaling of the
size distribution $\varpi_{G}(n)$ as
$\langle n\rangle^{-1} f(n/\langle n\rangle)$ leads to good collapse.
So, agreement with the results of the molecular simulations is much
more satisfactory than was the case for the E model.

\begin{figure}
\psfrag{epsilon}{{\large $\epsilon/\epsilon_c$}}
\psfrag{figa}{{(a)}}
\psfrag{figb}{{(b)}}
\includegraphics[width=0.45\textwidth,clip]{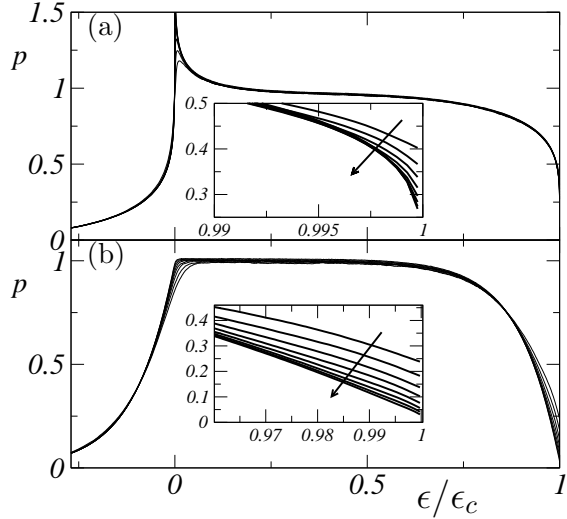}
\caption{\label{fig:p} Steady state distribution $p$ of zone strains vs $\epsilon/\epsilon_c$
for:
(a) E model; (b) G model. Inserts: blow-ups of the near-threshold region. 
The arrows indicate increasing values of $N=250\times2^q$ ($q=0,\ldots,7$).
}
\end{figure}

We now attempt to clarify how the noise spectrum affects so significantly the
avalanche behavior. We plot on Figure \ref{fig:p} the strain (ergodic)
averages of the distributions $p_{E,G}(\epsilon)$ of zone strains
in steady state for the two models and various system sizes $N$ ranging
from $250$ to $32000$. For each model, $p$ converges rapidly almost
everywhere toward a limit curve. This reflects into a very weak size
dependence of the macroscopic stress $\bar{\sigma} = 2 \mu \langle
\epsilon_{i} \rangle$, which varies by less than $0.4 \%$ (G model)
and $0.02 \%$ (E model) when $N$ increases from $1000$ to $32000$.

While $p_{E}$ and $p_{G}$ are similar in most of the $\epsilon$ range,
they present significant differences in the two regions $\epsilon
\sim 0$ and $\epsilon \sim \epsilon_{c}$. The peak in $p_{E}$ results
from refeeding zones at $\epsilon = 0$ after flips. The larger an
avalanche is, the more the corresponding peak is broadened by ulterior
flips within the avalanche itself. We therefore attribute the washing
out of the peak for the G model to the fact that it exhibits much
larger avalanches. More significant for our purpose is the detailed
behavior of $p$ near threshold, which reflects avalanche statistics.
In particular, $p_{c} = p(\epsilon_{c})$ is directly related to
$\langle n \rangle$. Indeed, the average flip and avalanche rates
verify $\mathcal{R}_{flip} = \langle n \rangle \mathcal{R}_{av}$. On
the other hand, in steady state, the flux of zones which cross
threhold under the effect of advective elastic loading, i.e. which
initiate avalanches, is $f = \mathcal{R}_{av} = \dot{\epsilon} p_{c} N$.
So:
\begin{equation}
\label{eq:pc}
p_{c} = \frac{1}{\langle n \rangle \epsilon_{c}}
\end{equation}
Since the data indicate that $\langle n \rangle$ diverges for $N
\rightarrow \infty$, they also indicate that the absorbing boundary
condition $p_{c} = 0$ should hold asymptotically here. Relation
(\ref{eq:pc}) provides a consistency test of our calculations. We
determine $p_{c}$ with the help of a second order polynomial
extrapolation near $\epsilon_{c}$, with sampling intervals $1.25
\times 10^{-6} \epsilon_{c}$. We find that relation (\ref{eq:pc})
holds within $1\% $ for the G model and $3\% $ for the E one.

In order to try and obtain analytical estimates for
avalanche size scalings, we describe the evolution of $p(\epsilon)$
by the approximate master equation :
\begin{equation}
\label{eq:master}
\frac{\partial p}{\partial t} = - \dot{\epsilon} \frac{\partial p}{\partial
\epsilon} + \int_{-\infty}^{\epsilon_{c}}\,d\epsilon' p(\epsilon')
w(\epsilon -\epsilon') - \Gamma p(\epsilon) + \frac{f}{N}
\delta(\epsilon)
\end{equation}
where $w$ is the single flip transition probability
\begin{equation}
\label{eq:transition}
w(\delta\epsilon) = N \frac{\dot{\epsilon}}{\epsilon_{c}}
\Pi(\delta\epsilon)
\end{equation}
with $½i$ the noise distribution, and $\Gamma =
\int_{-\infty}^{\infty}d(\delta\epsilon) w(\delta\epsilon)$. The delta
term, proportional to the normalized zone flux $f/N =
\dot{\epsilon}/\epsilon_{c}$, accounts for post-flip reinjection and
ensures the conservation of zone number.

In this approximation,
advection operates between all single flips, which amounts to
neglecting intra-avalanche time correlations. This leads to an
average advective $\epsilon$-shift during an avalanche
$\Delta\epsilon_{adv} \sim \mathcal{R}_{av} \sim \epsilon_{c}\langle
n\rangle/N$, to be compared with the average diffusive broadening
$\Delta\epsilon_{diff} \sim \sqrt{\langle n \rangle M_{2}(N)}$, with
$M_{2}(N) = \epsilon_{c}^{2}/N$ the noise variance. Hence,
$\Delta\epsilon_{adv}/\Delta\epsilon_{diff} \sim \sqrt{\langle n
\rangle}\sim N^{-(1-\alpha)/2}$, which suggests that our
approximation should improve in the large $N$ limit.

Integration of equation (\ref{eq:master}) in steady state yields:
\begin{equation}
\label{eq:master2}
\frac{f}{N} = \frac{\dot{\epsilon}}{\epsilon_{c}} = \dot{\epsilon}
p_{c} + \int_{-\infty}^{\epsilon_{c}}\,d\epsilon
p(\epsilon)\int_{\epsilon_{c}}^{\infty}\,d\epsilon' w(\epsilon -
\epsilon')
\end{equation}
Since $w$ is peaked around zero, in the spirit of the Fokker-Planck
approximation,we expand $p(\epsilon)$ close to
$\epsilon_{c}$ to first order: $p(\epsilon) \simeq p_{c} +
(\epsilon - \epsilon_{c})p'_{c}$. Using $p_{c} = \left(\langle n
\rangle\epsilon_{c}\right)^{-1}$, we obtain for the avalanche average
size:
\begin{equation}
\label{eq:anal}
\langle n \rangle = \left[ 1 + \frac{N
M_{1+}}{\epsilon_{c}}\right] \left[ 1 + \frac{N p'_{c}
M_{2+}}{2}\right]^{-1}
\end{equation}
where the (semi)-moments
\begin{equation}
\label{eq:semi}
M_{r+} = \int_{0}^{\infty}\,d(\delta\epsilon) w(\delta\epsilon)
\delta\epsilon^{r}
\end{equation}
For both models $M_{2+} = \epsilon_{c}^{2}/\pi^{2}N$, while
$M_{1+}^{(G)} =\epsilon_{c}/\pi\sqrt{\pi N}$ and $M_{1+}^{(E)}$ is given
by equation (\ref{eq:M1}).

If $p'_{c}$ converges towards a finite value ${p'}_{c}^{(\infty)}$,
equation (\ref{eq:anal}) predicts that, for large systems,

-- for the G model: \,\,\,$\langle n \rangle \sim N^{1/2}$

-- for the E model: \,\,\,$\langle n \rangle \sim \log N$

While this prediction accounts satisfactorily for the numerical
results for the gaussian model, we have checked (see also Figure
\ref{fig:sizeE}) that the log scaling is ruled our by our data. The
reason for this failure is illustrated by the insert of
Figure \ref{fig:p} (top). For the E model, we find that, for
increasing $N$, $p$ becomes increasingly steep in the near vicinity
of $\epsilon_{c}$. In the $N$-range investigated, we see a marked,
non-saturating, increase of $\mid p'_{c}\mid$ suggesting a possible
divergence, higher derivatives increasing even faster. This highly
singular behavior, reminiscent of that analyzed by Chabanol and Hakim~\cite{ChabanolHakim1997}
for a related model, clearly invalidates the above Fokker-Planck
expansion for model E. Conversely, for the G model, we find
numerically that $p'_{c}\epsilon_{c}^{2}/\pi^{2}$ does exhibit
convergence, towards $\simeq - 1.3$. This regular behavior suggests
that a Fokker-Planck expansion of the master equation
(\ref{eq:master}) should be valid for model G in the large $N$ limit.
We have indeed checked that, for increasing $N$, the steady state
distribution $p_{G}$  converges rapidly towards the solution of this
FP equation.

So, while even a schematic representation of long range elastic
couplings suffices to account for avalanches with power law size
scaling,
this discussion underscores that the detailed shape of the
noise spectrum is of crucial importance. Indeed, not only does it
affect the scaling exponent, but, as well, the scaling properties of
the distribution of avalanche sizes.\\

We consider that the results presented here, though still preliminary,
clearly show that the
dynamical noise due to long range elastic couplings is a key
ingredient that must be included in phenomenologies of plasticity of
amorphous solids. As $\dot{\gamma}$ increases beyond
the limit of the QS rgime, since avalanches are no longer separable,
the spectrum of the flip-generated dynamical noise will of course
change. The question of its evolution with $\dot{\gamma}$, as well as
that of its interplay with thermal noise at finite temperature remain for
the moment completely open ones.
Besides, the above discussion indicates two main
routes for further investigation.

On the one hand, a more realistic modelization of elastic couplings in
the presence of structural disorder is needed. Indeed, Leonforte {\it et al}~\cite{TanguyWittmerLeonforteBarrat2002,LeonforteTanguyWittmerBarrat2004}
have shown that the elastic response of amorphous solids self-averages
into the continuum elastic response only beyond a length scale $\xi$
of order $\sim 20$ atomic diameters $a_{0}$. For $r < \xi$, the
elastic response is dominated by non-affine effects. So, for the
existing molecular simulations focussing on avalanche dynamics, where
$L$ is limited to $\lesssim 50 a_{0}$, noise tails are very likely to
be controlled by elastic non-affinity. This issue, as well as that of
the statistical distribution of zone parameters (such as shear
modulus and threshold strain), will demand the development of a
coarse-grained description of the elasticity of glassy systems.

On the other hand, our mean-field approximation wipes out from the
start the correlation anisotropy arising from the quadrupolar symmetry
of elementary events, responsible for the preferential
avalanche orientations observed in the glass simulations of Maloney et
al~\cite{MaloneyLemaitre2006} and Tanguy et al~\cite{TanguyLeonforteBarrat2006}. 
In order to evaluate the robustness of the
mean-field scalings and also start addressing the issue of
localization, full simulations of  model E in the presence of rigid
boundaries will be necessary.\\

{\it Acknowledgements}

We are grateful to V. Hakim and F. Lequeux for useful comments.


\begin{thebibliography}{10}

\bibitem{FalkLanger1998}
M.~L. Falk and J.~S. Langer, Phys. Rev. E {\bf 57},  7192  (1998).

\bibitem{Sollich1998}
P. Sollich, Phys. Rev. E {\bf 58},  738  (1998).

\bibitem{MaloneyLemaitre2006}
C. Maloney and A. Lema\^{\i}tre, Phys. Rev. E {\bf 74},  016118  (2006).

\bibitem{MaloneyLemaitre2004}
C. Maloney and A. Lema\^{\i}tre, Phys. Rev. Lett. {\bf 93},  16001  (2004).

\bibitem{BaileySchiotzLemaitreJacobsen2007}
N.~P. Bailey, J. Schiotz, A. Lemaitre, and K.~W. Jacobsen, Phys. Rev. Lett.
  {\bf 98},  095501  (2007).

\bibitem{Eshelby1957}
J.~D. Eshelby, Proc. Roy. Soc. London A {\bf 241},  376   (1957).

\bibitem{LemaitreCaroli2007}
A. Lemaitre and C. Caroli, cond-mat/0705.0823, 2007.

\bibitem{ArgonBulatovMottSuter1995}
A.~S. Argon, V.~V. Bulatov, P.~H. Mott, and U.~W. Suter, J. Rheol. {\bf 39},
  377  (1995).

\bibitem{PicardAjdariLequeuxBocquet2004}
G. Picard, A. Ajdari, F. Lequeux, and L. Bocquet, Eur. Phys. J. E {\bf 15},
  371  (2004).

\bibitem{footnote}
Clearly, such a statistical description of avalanches also demands
that $\langle n \rangle \ll N$. We will see that this condition is fulfilled in our
model.

\bibitem{ChenBakObukhov1991}
K. Chen, P. Bak, and S.~P. Obukhov, Phys. Rev. A {\bf 43},  625  (1991).

\bibitem{MaloneyLemaitre2004a}
C. Maloney and A. Lema\^{\i}tre, Phys. Rev. Lett. {\bf 93},  195501  (2004).

\bibitem{LeonforteTanguyWittmerBarrat2004}
F. Leonforte, A. Tanguy, J.~P. Wittmer, and J.-L. Barrat, Phys. Rev. B {\bf
  70},  014203  (2004).

\bibitem{ChabanolHakim1997}
M.~L. Chabanol and V. Hakim, Phys. Rev. E {\bf 56},  R2343  (1997).

\bibitem{TanguyWittmerLeonforteBarrat2002}
A. Tanguy, J.~P. Wittmer, F. Leonforte, and J.-L. Barrat, Phys. Rev. B {\bf
  66},  174205  (2002).

\bibitem{TanguyLeonforteBarrat2006}
A. Tanguy, F. Leonforte, and J.~L. Barrat, Eur. Phys. J. E {\bf 20},  355
  (2006).

\end{thebibliography}


%

\end{document}